# Optimization of a traveling wave superconducting radiofrequency cavity for upgrading the International Linear Collider

V. Shemelin, H. Padamsee, V. Yakovlev

*FNAL, Batavia, IL 60510, U.S.A.*

The Standing Wave (SW) TESLA niobium-based superconducting radio frequency structure is limited to an accelerating gradient of about 50 MV/m by the critical RF magnetic field. To break through this barrier, we explore the option of niobium-based traveling wave (TW) structures. Optimization of TW structures was done considering experimentally known limiting electric and magnetic fields. It is shown that a TW structure can have an accelerating gradient above 70 MeV/m that is about 1.5 times higher than contemporary standing wave structures with the same critical magnetic field. The other benefit of TW structures shown is R/Q about 2 times higher than TESLA structure that reduces the dynamic heat load by a factor of 2. A method is proposed how to make TW structures multipactor-free. Some design proposals are offered to facilitate fabrication. Further increase of the real-estate gradient (equivalent to 80 MV/m active gradient) is also possible by increasing the length of the accelerating structure because of higher group velocity and cell-to-cell coupling. Realization of this work opens paths to ILC energy upgrades beyond 1 TeV to 3 TeV in competition with CLIC. The paper will discuss corresponding opportunities and challenges.

## I. INTRODUCTION

A strong physics attraction for the ILC - besides the Higgs and Top Factories [1-2] - is the inherent energy upgradability. As described in the ILC TDR [3], ILC offers paths to energy upgrades of 0.5 TeV and 1 TeV for which higher gradients are critical for affordability, as cavities and cryomodules are dominant cost drivers. There has been steady progress in single and multicell cavity gradients [4] over the last 3+ decades along with SRF science and technology advances. Proof-of-principle is already in hand for cavity preparations that deliver single cell TESLA-shape cavities with gradients up to 49 MV/m [5-6], and for 9-cell cavities with gradients up to 45 MV/m [7]. These gradient advances come from high purity, high RRR Nb, electropolishing and optimized 120°C baking, 800°C furnace treatment for hydrogen removal, and 100 atm high pressure water rinsing for removal of field emission particulates. The fundamental critical magnetic field of approximately 210 mT presents the ultimate hard limit to niobium cavity gradients. For the standing wave TESLA shape structure, with *peak surface magnetic field to accelerating field ratio* $B_{pk}/E_{acc}$ = 4.26 mT/(MV/m), this limit translates to a maximum gradient of 50 MV/m. The peak electric field also presents a limit due to field emission, but this is a practical – not fundamental - limit which in principle can be overcome with technology advances in surface preparation (such as more effective final high-pressure rinsing). Further gradient advances from 50 – 59 MV/m have been demonstrated [8-10] with single cell cavities of advanced cavity geometries with 10 – 15% lower $B_{pk}/E_{acc}$, such as Re-entrant, Low Loss, Ichiro, and Low Surface Fields shapes.

Even higher gradients are needed for ILC energy upgrades beyond 1 TeV! This paper discusses optimized *traveling wave* (TW) superconducting niobium-based structures [12] with effective gradients up to 73 MV/m to open upgrade paths to 3 TeV, in competition with CLIC at 3 TeV. A previous report [11] shows the overall cost for 3 TeV ILC with 70 MV/m gradient is comparable ($1B lower) to CLIC 3 TeV, and the AC power is 80 MW lower. TW structures offer two main advantages compared to standing wave (SW) structures: substantially lower peak magnetic and peak electric field ratios, and substantially higher *R/Q* (for lower cryogenic losses). In addition, TW structure operates far of the passband boundaries, and therefore, has high stability of the field distribution along the structure with respect to geometrical perturbations [12]. This allows a much longer structure length and hence no gaps between short cavities, thereby increasing the real-estate gradient, but this advantage substantially increases the engineering challenges. Besides, the TW structure requires a feedback waveguide for redirecting power from the end of the structure back to the front to avoid high peak surface fields in the accelerating cells. This requires careful tuning to compensate reflections along the TW ring to obtain a pure traveling wave regime at the desired frequency. Because the beam bunch charge for the 3 TeV upgrade is 3 times lower than the bunch charge for 0.5 TeV [11], it is further possible to lower the aperture (from 70 mm to 50 mm) to obtain an overall 50 % reduction in $B_{pk}/E_{acc}$ and factor of 2 gain in *R/Q* over the TESLA standing wave structure.

Previously, a substantial progress was made at Fermilab and in Euclid Techlab on the way to realization of the TW structure in the regime of a resonant ring [12 - 15]. The present work makes use of that progress to advance the topic further.

The optimizations described below are enabled by accurate calculations of cavity parameters. 2D computer code SuperLANS [16] has the accuracy necessary for these optimizations.

## II. GEOMETRY OF AN ELLIPTICAL CAVITY

Contemporary superconducting RF cavities for high energy particle accelerators consist of a row of cells coupled together as shown in Fig. 1a. The contour of a half-cell consists of two elliptic arcs and a straight segment tangential to both. The contour can be described by several geometrical parameters shown in Fig. 1b. Three of these parameters, length of the half-cell $L$, aperture $R_a$, and equatorial radius $R_{eq}$ are defined by physical requirements: in the case of a traveling wave $L = \theta\lambda/(2\pi)$, where $\theta$ is the phase advance per cell, and $\lambda$ is the wave length; the aperture is defined by requirements for coupling between cells and by the level of wake fields that can be allowed for a given accelerator; and the equatorial radius $R_{eq}$ is used for tuning the cavity to a given frequency. Remaining four parameters ($A$, $B$, $a$, $b$) can fully describe the geometry. Here $A$, $B$ and $a$, $b$ are the half-axes of the equatorial and iris constitutive ellipses, respectively. The best combination of four parameters is the goal for the cavity shape optimization. The angle of the wall inclination between the axis of rotation and the straight segment of the wall is designated as $\alpha$. The cavity with $\alpha < 90°$ is known as the reentrant cavity.

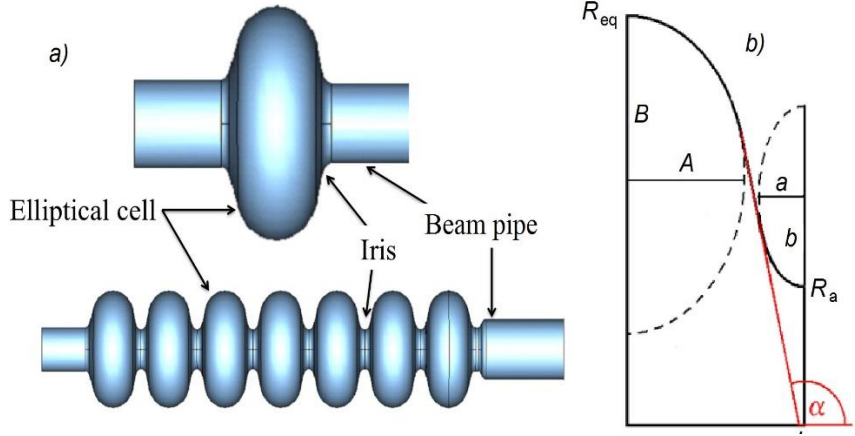

Figure 1. a) Single cell and multicell elliptical cavities; b) geometry of the half-cell.

### III. CELLS WITH DIFFERENT $L - A$

Theoretically, the value of $A$ (see Fig. 1b) is limited by the value of $L$ though there is a technological limit for the gap between outer surfaces of cells, $2(L-A-t)$, defined by the thickness $t$ of material and needed for welding two half-cells together.

To investigate the whole set of parameters, we will optimize the cavity with $E_{pk}/E_{acc} = 2$ for different values of $L-A$. Here, the phase advance angle $\theta = 105°$ and the aperture radius $R_a = 30$ mm. The results for $L - A = 0$, 1, and 5 are presented in Table I.

Table I. Parameters of optimized cells for different values of $L - A$. $E_{pk}/E_{acc} = 2$, phase advance per cell $\theta = 105°$, aperture $2R_a = 60$ mm.

| $L$-$A$, mm | 0 | 1 | 5 |
|---|---|---|---|
| $A$, mm | 33.631 | 32.631 | 28.631 |
| $B$, mm | 34.294 | 37.304 | 38.919 |
| $a$, mm | 5.284 | 5.233 | 4.903 |
| $b$, mm | 7.777 | 7.680 | 6.790 |
| $B_{pk}/E_{acc}$, mT/(MV/m) | 2.848 | 2.833 | 3.027 |
| $R_{sh}/Q$, Ohm/m | 1995 | 1967 | 1820 |
| $\alpha$, degrees | 71.81 | 73.26 | 90.23 |
| $R_{eq}$, mm | 95.526 | 96.919 | 100.255 |

One can see from the Table that benefits of the reentrant shape, lower $B_{pk}/E_{acc}$ and higher $R_{sh}/Q$, cannot be realized for the realistic cavity with $\theta = 105°$ because of shorter cell length than in the 180° standing wave cavity, and the necessity to have a non-zero thickness of material and the gap for welding are more crucial than for the 180° cavity.

A discussion with an expert [17] revealed that even the value of $L-A = 5$ mm, is hardly possible, but we found a solution implemented by AES, Fig. 2, with a diaphragm thickness of 11 mm ($2 \times (1.348 − 1.131) \times 25.4 = 11.024$), so, at least $L − A = 5.5$ mm can be done.

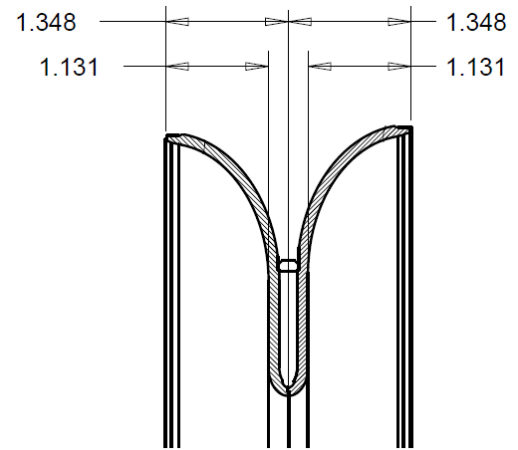

Figure 2. An example of cavity with $L$ - $A$ =5.5 mm.

A possible solution of this problem can be changing of the cavity design, see Fig. 3. Here, instead of two half-cells welded together, the half-cells are welded to an iris disc. The place of the weld from the inner side of the cells can be taken at the circle where the electric and magnetic fields are in a balance $\varepsilon_0 E^2 = \mu_0 H^2$. In this case, both fields are much lower than at their maxima, and a small perturbation by the welding bead will not change the frequency and other important figures of merit. This design makes possible to get rid of the stiffening rings because the iris disc is stiff. Somewhat increased distance from the iris tip to the cooling agent is not important because heat production in this area is negligible. The iris tip can be made of any shape in accordance with optimization. The radius of curvature at the iris tip becomes comparable to the thickness of the niobium sheet when the shape is optimized, that makes difficult to make it by deep drawing. In the case of a solid disc, this problem is removed. To decrease the cost of the iris disc, it can be made from a lower RRR niobium.

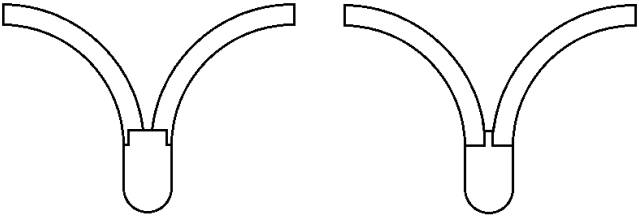

Figure 3. New possible design of the cells' connection.

For production of the iris disc, the experience of production TW X-band normal conducting accelerating structures can be used [16]. This structure reached the *accelerating gradient up to 100 MeV/m* that was due not only to higher frequency but also to diamond turning which gives very high smoothness of surface and precise dimensions as compared to deep drawing. Cavities machined out of bulk niobium are expensive [20], but the iris disc for a TW cavity will be much thinner and smaller in diameter than we would use to make a whole SW cavity. Combination of diamond turning, and chemical treatment and rinsing used for SW cavities can further improve the surface quality needed for suppression of field emission. In optimization, the higher is $E_{pk}/E_{acc}$, the lower $B_{pk}/E_{acc}$ can be reached as it was shown in [21].

### IV. CELLS WITH DIFFERENT APERTURES $2R_a$

Smaller aperture dramatically increases the accelerating rate because of smaller $B_{pk}/E_{acc}$ for the same value of $E_{pk}/E_{acc}$ and, hence lower probability of magnetic quench. This decrease of $B_{pk}/E_{acc}$ together with increase of $R_{sh}/Q$ vs. decrease of $R_a$ is illustrated in Table II.

### V. NEW APPROACH FOR OPTIMIZATION

Optimization of an elliptical cavity is usually done as a search for minimum $B_{pk}/E_{acc}$ when the value of $E_{pk}/E_{acc}$ is given. It is also possible to minimize $E_{pk}/E_{acc}$ for a given $B_{pk}/E_{acc}$ but the truth is that we need to reach as high as possible accelerating gradient $E_{acc}$ before field emission or magnetic quench limit further increase of the accelerating gradient. So, the ideal situation would be to reach both limits simultaneously using all the possibilities to increase $E_{acc}$. If we know the maximal achievable surface peak fields $E_{pk}^*$ and $B_{pk}^*$, then the cavity having equal values of $E_{pk}/E_{pk}^*$ and $B_{pk}/B_{pk}^*$ will be at equal distances from either limit. Then the criterion of the shape optimization can be written as minmax ($E_{pk}/E_{pk}^*$, $B_{pk}/B_{pk}^*$).

Table II. Parameters of optimized cells for different values of aperture radius $R_a$. $E_{pk}/E_{acc}$ = 2, phase advance per cell $\theta = 105°$, aperture $2R_a$ = 60 mm.

| $R$a, mm | 35 | 30 | 25 |
|---|---|---|---|
| $A$, mm | 28.631 | 28.631 | 28.631 |
| $B$, mm | 39 | 38.919 | 36.986 |
| $a$, mm | 6.2 | 4.903 | 3.872 |
| $b$, mm | 8.7 | 6.790 | 5.159 |
| $B_{pk}/E_{acc}$, mT/(MV/m) | 3.219 | 3.027 | 2.858 |
| $R_{sh}/Q$, Ohm/m | 1607 | 1820 | 2.048 |
| $\alpha$, degrees | 85.77 | 90.23 | 91.98 |
| $R_{eq}$, mm | 101.333 | 100.255 | 99.035 |

In the optimization, absolute values of $E_{pk}$ and $B_{pk}$ do not matter, important is only their ratio. Absolute values depend on normalization used in the program for the cavity calculation, e.g., the total electromagnetic energy stored in the cavity or the accelerating rate. Their ratio depends on the geometry only. Values under the sign of minmax (see above) become equal in the result because $E_{pk}$ and $B_{pk}$ change reversely: when one of them increases, the other decreases, and vice versa.

The same statement about ratio is true for $E_{pk}^*$ and $B_{pk}^*$: optimization for $E_{pk}^*$= 120 MV/m and $B_{pk}^*$= 240 mT will be the same as optimization for $E_{pk}^*$= 100 MV/m and $B_{pk}^*$= 200 mT. We can consider the first pair of parameters as the aggressive version for the future cavities, and the second pair as a basic version. In this case, we need to do only one optimization for both cases. Let us call this optimization "optimization 100/200".

Some experts [18] recommend this and even lower, 70 MV/m, limit electric field, with the same magnetic field limit, 200 mT, possibly because of difficulties to provide high quality of surface preparation in a multicell cavity.

A possible future progress in the increase of achievable fields can change this proportion, and we have this proportion changed [10]: a gradient of 59 MV/m was achieved in a single-cell cavity that corresponds to a peak surface electric field of 125 MV/m and a peak magnetic field of 206.5 mT. The gradient was limited by a hard quench. We will make another optimization with these parameters 125/206.5 ≈ 120/200 and call it "optimization 120/200". The comprehension that quench governed by $B_{pk}$, is a hard limit, whereas field emission, governed by $E_{pk}$, could be decreased with better cleaning made possible to achieve this record gradient.

Optimization for minimum $B_{pk}/E_{acc}$ when the value of $E_{pk}/E_{acc}$ is given, can be revised in the light of the method proposed here. For example, well optimized for a given aperture, the TESLA cavity has $E_{pk}/E_{acc} = 2$ and $B_{pk}/E_{acc} = 4.2$ mT/(MV/m). If we assume that both limits, $E^*_{pk}$ and $B^*_{pk}$, are achieved simultaneously in this optimization, then $E_{pk}/B_{pk} = E^*_{pk}/B^*_{pk} = 2/4.2$ (MV/m)/mT $= 100/210$ (MV/m)/mT. This means that this cavity can be treated as a cavity optimized for $E^*_{pk} = 100$ MV/m and $B^*_{pk} = 210$ mT, or, proportionally, for example, for $E^*_{pk} = 80$ MV/m and $B^*_{pk} = 168$ mT.

The difference between these two methods is in the fact that we do not know *a priori* what value of $B_{pk}/E_{acc}$ we will have for a given value of $E_{pk}/E_{acc}$ in the old method, but in the new method, we can choose the ratio between the extremal fields based on experiment, and then perform the optimization.

## VI. FIRST OPTIMIZATIONS 100/200 AND 120/200 FOR A TRAVELING WAVE. APERTURE $2R_a = 50$ mm

Let us use the approach described above for optimization of a cell for a periodic structure with an aperture $2R_a = 50$ mm. We assume the limiting surface fields as (1) $E_{pk^*} = 100$ MV/m and $B_{pk^*} = 200$ mT, and (2) $E_{pk^*} = 120$ MV/m and $B_{pk^*} = 200$ mT. The phase advance is $\theta = 105°$ for the case (1) and varies from 90° to 110° for the case (2).

We will omit here details of the optimization, and present results only, see Table III and Fig. 4. A value of $E_{pk}/E_{acc}$ is low, as compared with standing wave cavities; possibly, this is an effect of transition to TW. The values of $R_{sh}/Q$, responsible for the dynamic heat load, is about 2 times higher in all further optimizations (Tables IV and V) of the TW structures than in the SW. Comparison with the TESLA cavity is presented in Table VI.

Table III. Parameters of optimized cells with limiting surface fields: (1) $E^*_{pk} = 100$ MV/m and $B^*_{pk} = 200$ mT, and (2) $E^*_{pk} = 120$ MV/m and $B^*_{pk} = 200$ mT; $L - A = 5$ mm, aperture radius $R_a = 25$ mm. $E^*_{acc}$ is the accelerating rate when the limiting surface fields are achieved.

| Optimization | 100/200 | 120/200 | 120/200 | 120/200 | 120/200 | 120/200 |
|---|---|---|---|---|---|---|
| Phase advance $\theta$, deg | 105 | 90 | 95 | 100 | 105 | 110 |
| $A$, mm | 28.631 | 23.826 | 25.428 | 27.029 | 28.631 | 30.232 |
| $B$, mm | 97.44 | 36.4 | 38.1 | 39.9 | 40.91 | 42.1 |
| $a$, mm | 6.084 | 4.512 | 4.840 | 5.171 | 5.494 | 5.817 |
| $b$, mm | 11.098 | 7.52 | 8.136 | 8.772 | 9.379 | 9.986 |
| $E_{pk}/E_{acc}$ | 1.655 | 1.727 | 1.730 | 1.734 | 1.739 | 1.745 |
| $B_{pk}/E_{acc}$, mT/(MV/m) | 3.309 | 2.878 | 2.883 | 2.890 | 2.898 | 2.909 |
| $R_{sh}/Q$, Ohm/m | 1789 | 2127 | 2096 | 2063 | 2029 | 1992 |
| $\alpha$, degrees | 94.73 | 90.91 | 90.33 | 89.61 | 88.77 | 87.71 |
| $R_{eq}$, mm | 106.156 | 98.950 | 98.991 | 99.068 | 99.016 | 99.011 |
| $v_{gr}/c$ | 0.01365 | 0.01831 | 0.01776 | 0.01710 | 0.01635 | 0.01551 |
| $E^*_{acc}$, MV/m | 60.4 | 69.5 | 69.4 | 69.2 | 69.0 | 68.8 |
| $E^*_{acc} \times 2L$, MV | 4.06 | 4.00 | 4.22 | 4.43 | 4.64 | 4.85 |

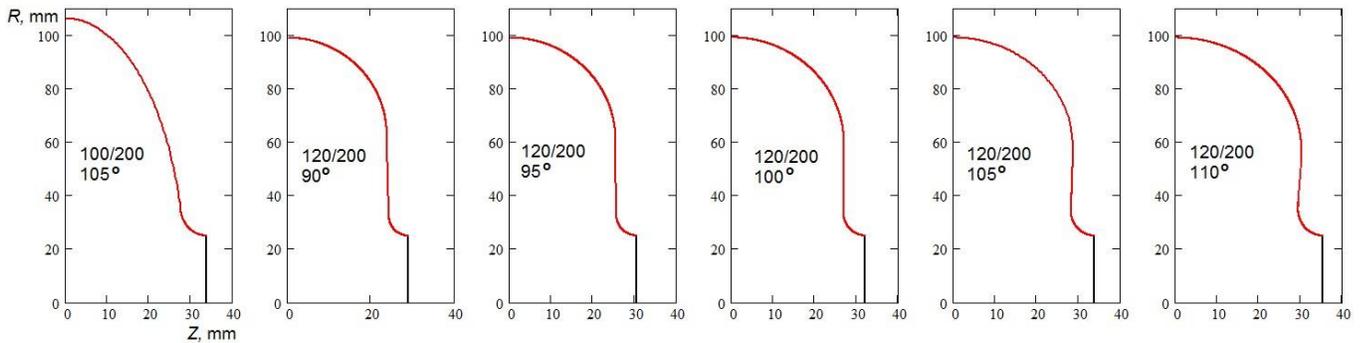

Figure. 4. Cells with parameters presented in Table III.

All most important parameters – $E_{pk}/E_{acc}$, $B_{pk}/E_{acc}$, $R_{sh}/Q$, $E^*_{acc}$ – are improving when the phase advance per cell $\theta$ decreases. However, acceleration per cell, $E^*_{acc} \times 2L$, reduces, and the shorter the cell the less acceleration it provides ($2L$ is the cell length). Because of high cost of superconducting cell fabrication, the number of cells should be minimized. It can also be shown that requirements to the accuracy of dimensions are more stringent when the dimensions defining the cell's shape decrease. So, the trade-off for the phase advance per cell should be defined. A possible solution is a cell with the wall slope angle $\alpha = 90°$ that correspond to $\theta$ between 95° and 100°. Cells with $\alpha = 90°$ seem simpler for manufacturing and chemical treatment compared to reentrant cells.

## VII. OPTIMIZATIONS 120/200 FOR A TRAVELING WAVE. APERTURE $2R_a$ = 40 mm

To investigate benefits of a smaller aperture, data for a cavity with $R_a$ = 20 mm are presented in Table IV. Phase advance of 90° is chosen; as can be seen from Table III, dependence of $E^*_{acc}$ on this value is not strong.

Table IV. Parameters of an optimized cell with the limiting surface fields $E^*_{pk}$ = 120 MV/m and $B^*_{pk}$ = 200 mT; $L - A$ = 5 mm, aperture radius $R_a$ = 20 mm. $E^*_{acc}$ is the accelerating rate when the limiting surface fields are achieved.

| Optimization | 120/200 |
|---|---|
| Phase advance $\theta$, deg | 90 |
| $A$, mm | 23.826 |
| $B$, mm | 35 |
| $a$, mm | 3.874 |
| $b$, mm | 6.777 |
| $E_{pk}/E_{acc}$ | 1.639 |
| $B_{pk}/E_{acc}$, mT/(MV/m) | 2.732 |
| $R_{sh}/Q$, Ohm/m | 2367 |
| $\alpha$, degrees | 91.74 |
| $R_{eq}$, mm | 97.990 |
| $v_{gr}/c$ | 0.009315 |
| $E^*_{acc}$, MV/m | 73.2 |
| $E^*_{acc} \times 2L$, MV | 4.22 |

## VIII. COMPARISON OF OPTIMIZATIONS 120/200, 150/200, AND 200/200. APERTURE $2R_a$ = 50 mm, WALL SLOPE ANGLE $\alpha$ = 90°, PHASE ADVANCE $\theta$ = 100°, $a$ = 5 mm

To understand benefits of further improvement of the surface for decreasing the dark currents, optimization was done for a simplified geometry with $A + a = L$, where $a = 5$ mm, i.e. for the wall slope angle $\alpha = 90°$. The phase advance was taken $\theta = 100°$ because in previous optimization this phase advance led to $\alpha$ close to 90°. Results are presented in Table V.

One can see that an increase in the limiting field $E^*_{pk}$ from 120 to 150 MV/m, i.e., by 25 %, leads to increase in the acceleration rate $E^*_{acc}$ by 1.8 % only, and an increase from 150 to 200 MV/m, i.e., by 33 %, leads to increase in the $E^*_{acc}$ by 0.4 %. So, it does not make any sense to hope for increase in accelerating rate if the limiting electric field will be above 120 MV/m. The main limitation is the surface magnetic field.

Table V. Parameters of optimized cells with limiting surface fields (1) $E^*_{pk}$ = 120 MV/m and $B^*_{pk}$ = 200 mT, (2) $E^*_{pk}$ = 150 MV/m and $B^*_{pk}$ = 200 mT, and (3) $E^*_{pk}$ = 200 MV/m and $B$ = 200 mT; $a = L - A$ = 5 mm, aperture radius $R_a$ = 25 mm. $E^*_{acc}$ is the accelerating rate when the limiting surface fields are achieved.

| Optimization | 120/200 | 150/200 | 200/200 |
|---|---|---|---|
| Phase advance $\theta$, deg | 100 | 100 | 100 |
| $B$, mm | 51.5 | 39 | 39 |
| $b$, mm | 8.258 | 4.200 | 2.086 |
| $E_{pk}/E_{acc}$ | 1.752 | 2.152 | 2.857 |
| $B_{pk}/E_{acc}$, mT/(MV/m) | 2.919 | 2.870 | 2.857 |
| $R_{sh}/Q$, Ohm/m | 2029 | 2095 | 2117 |
| $R_{eq}$, mm | 101.192 | 98.730 | 98.569 |
| $v_{gr}/c$ | 0.01695 | 0.01595 | 0.01499 |
| $E^*_{acc}$, MV/m | 68.5 | 69.7 | 70.0 |
| $E^*_{acc} \times 2L$, MV | 4.39 | 4.46 | 4.48 |

## IX. MULTIPACTOR CONSIDERATION

According to [22], existence of multipactor in a cavity is defined by the geometrical parameter $p$: experimental data presented in this book show that at $p = 0.3$ and higher there is strong multipactor. The TESLA shape cavity has $p = 0.286$ and has weak multipactor activity. The Cornell ERL cavity has $p = 0.276$, and multipactor in it was not observed. With some degree of caution, one can say that $p = 0.28$ is a safe limit for multipactor. The case in point is the elliptical niobium superconducting cavity with a standard treatment of the surface, other materials can give different limits of $p$. We define $p$ for a two-point multipactor to which the elliptical cavities are susceptible foremost. One-point multipactor occurs if there is a long flat region in the cavity equator area that is not our case.

In TW regime, in distinction to SW, the maximum of $E$ or $H$ field is reached at different time for different points. For this reason, the fields can be presented in the form, e.g., $H(r,z,t) = U(r,z,t) + iV(r,z,t)$, where functions $U$ and $V$ reach their maxima with a shift in phase of 90°. Fortunately, the function $V$ is small compared to $U$ in the equator area, $L = 0$ in Fig. 5, and all the theory appears quite applicable to TW regime as it is for the SW regime.

If we calculate $p$ for cavities presented above, we will see that this parameter is higher than 0.28 and multipactor in some cases is inevitable. A solution of this problem can be a change of the cell shape that does not compromise the optimized parameters too much (say, not more than 1 % for $E_{pk}/E_{acc}$ and $B_{pk}/E_{acc}$) but decreases $p$ to a safe value. A possible way to decrease $p$ is in increase of the curvature radius at the cavity equator (formula (8.16) in [22]). A change in the iris half-axes $a$ and $b$ has a small influence on the value of $p$ and leads to a fast change in the value of $E_{pk}$, the value of half-axis $A$ is chosen as big as possible and cannot be increased (technological limitations) to decrease $B_{pk}$. So, we will decrease the length of the half-axis $B$ that is perpendicular to the cavity axis, and so increase the curvature radius at the equator $R_c = A^2/B$.

As an example, let us consider the cavity with $\theta = 100°$, $R_a = 25$ mm, optimized for "120/200", see Table III. Its original $p = 0.303$. When $B$ is decreased from 39.9 mm to 30 mm, $E_{pk}/E_{acc} = 1.734$ will remain practically the same with a slight increase of 0.013 %, $B_{pk}/E_{acc} = 2.890$ will increase to 2.913 mT/(MV/m), i.e., by 0.81 %. To keep the frequency, $f = 1300$ MHz, the equatorial radius will change from 99.068 to 97.144 mm. The new value of $p$ will be 0.279.

The relative change of $E_{pk}/E_{acc}$ and $B_{pk}/E_{acc}$ is designated in the Table VI as $\delta[E_{pk}/E_{acc}]$ and $\delta[B_{pk}/E_{acc}]$, in percent. The accelerating rate when one of the limiting surface fields is achieved, $E^*_{acc}$, will also decrease relative to the original value, but, again, not more than 1 %.

Such a transformation was done for several cell shapes from Table III. Results are presented in Table VI.

A segment of cavity consisting of 16 cells with dimensions from the first column of Table VI is shown in Fig. 6. Inner dimensions only are shown.

So, here the method of anti-multipactoral transformation is presented, and hopefully any shape of a cell chosen in the future can be transformed into a multipactor-free shape without big losses in the acceleration rate.

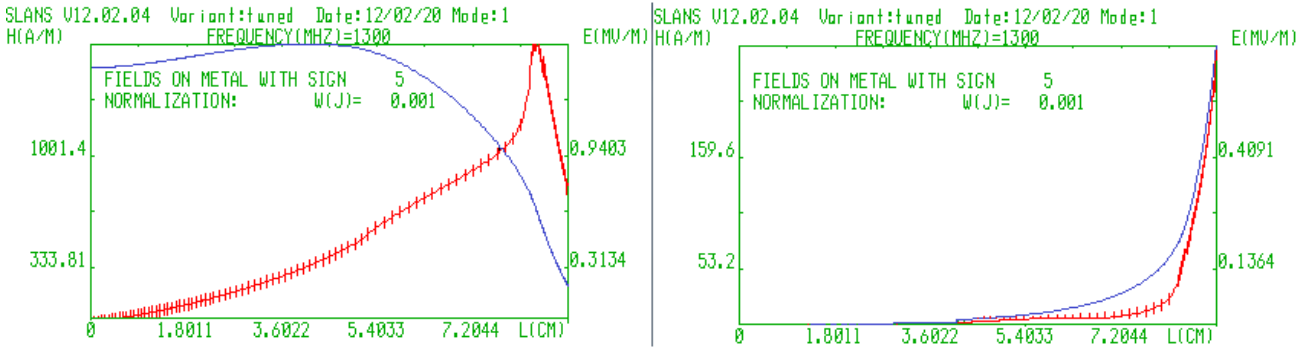

Figure 5. $U$ (left picture) and $V$ (right) components of the electric (red) and magnetic (blue) surface fields along the profile line for the geometry from the second column of Table V.

Table VI. Parameters of some TW cells from Table III before and after transformation removing multipactor conditions. Right column – parameters of the TESLA cavity, for comparison.

| Optimization | 120/200 | 120/200 | 120/200 | TESLA SW (100/210) |
|---|---|---|---|---|
| Phase advance $\theta$, degrees | 90 | 95 | 100 | 180 |
| | before/after | before/after | before/after | |
| $A$, mm | 23.826 | 25.428 | 27.029 | 42 |
| $B$, mm | 36.4/24 | 38.1/28 | 39.9/30 | 42 |
| $a$, mm | 4.512 | 4.840 | 5.171 | 12 |
| $b$, mm | 7.52 | 8.136 | 8.772 | 19 |
| $E_{pk}/E_{acc}$ | 1.727/1.728 | 1.730/1.730 | 1.734/1.734 | 1.99 |
| $\delta[E_{pk}/E_{acc}]$, % | 0.081 | 0.010 | 0.013 | |
| $B_{pk}/E_{acc}$, mT/(MV/m) | 2.878/2.897 | 2.883/2.904 | 2.890/2.913 | 4.16 |
| $\delta[B_{pk}/E_{acc}]$, % | 0.651 | 0.734 | 0.809 | |
| $R_{sh}/Q$, Ohm/m | 2127/2151 | 2096/2115 | 2063/2081 | 993 |
| $R_{eq}$, mm | 98.950/96.458 | 98.991/97.002 | 98.569/97.144 | 103.35 |
| Aperture radius $R_a$, mm | 25 | 25 | 25 | 35 |
| $p$ | **0.302/0.270** | **0.301/0.278** | **0.303/0.279** | 0.286 |

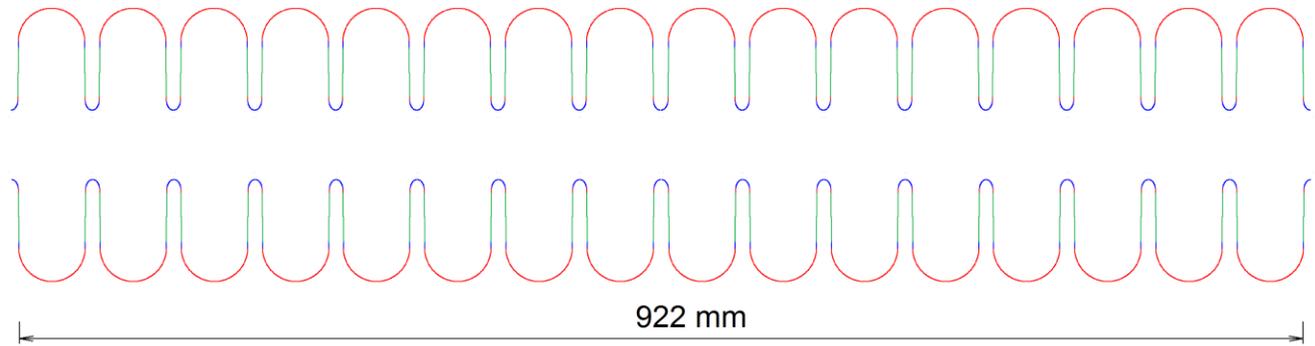

Figure 6. A segment of a TW cavity with dimensions from Table VI, column 1, after transformation.

## X. A POSSIBLE WAY TO FUTHER IMPROVE $B_{pk}/E_{acc}$

In optimization of a SW cavity cell, the magnetic field vs. the profile length has initially a maximum that should be flattened ([21], Fig. 5) to minimize $B_{pk}/E_{acc}$. It was expected that in the elliptic cell of a TW cavity the situation would be the same. An example of power density along the profile line of a TW cavity cell, which is proportional to the $B_{pk}^2$ is shown in Fig. 7.

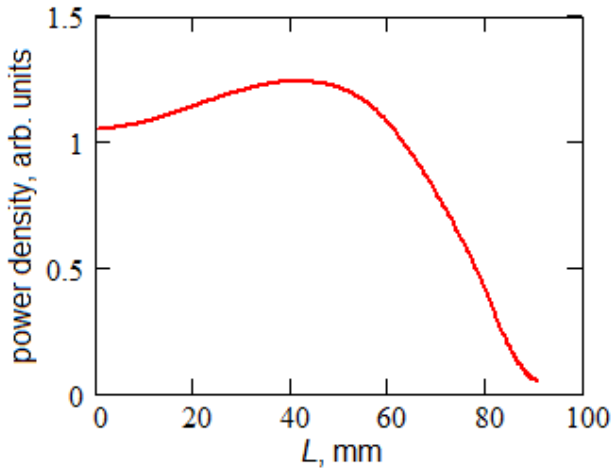

Figure 7. Power density distribution along the profile line of a TW cavity cell.

A possible way to solve this problem is to substitute the elliptic arc with another smooth curve. In Fig. 8 the elliptic arc with half-axes *A* and *B* is substituted with a curve consisting of 3 conjugated circular arcs.

An analogous procedure can be performed for the lower elliptic arc, defining the iris. This procedure will only slightly change $B_{pk}$ but can increase $E_{acc}$, so the ratio $B_{pk}/E_{acc}$ will be improved. It seems that there is a resource in the lower ellipse shape because the peak of electric field is very big – Fig. 5. If this peak will be shifted to the iris tip, the force lines will go closer to the axis increasing $E_{acc}$. But actually, both components of the electric field, *U* and *V*, should be taken into account, and the maximal field can be presented as $\sqrt{U^2+V^2}$, in this case the field near the iris tip will be significantly more flat.

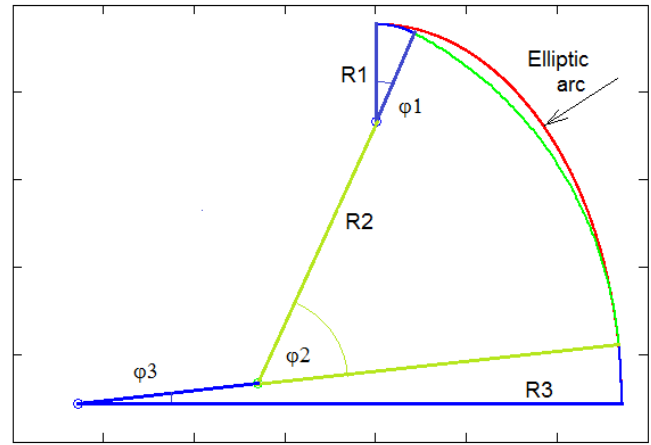

Figure 8. Substitution of an elliptic arc with three circular arcs.

This substitution of both upper (equatorial) and lower (iris) elliptic arcs with 3 circular arcs and fitting by trial and error for the case of a cell in the second column of Table V improved the value of $B_{pk}/E_{acc}$ from 2.870 to 2.831, i.e. by 1.4 % without changing $E_{pk}/E_{acc}$. However, because of a smaller radius of curvature at the equator, Fig. 9, the geometric parameter *p* increased from 0.299 to 0.318, making possible multipactor in this geometry.

This example shows a very big possibility of the shape modification with close values of important parameters. The addition of 6 circular arcs adds 6 degrees of freedom and optimization with these possibilities can be done if a special software is developed, taking into account some restriction, e.g., the value of *p*.

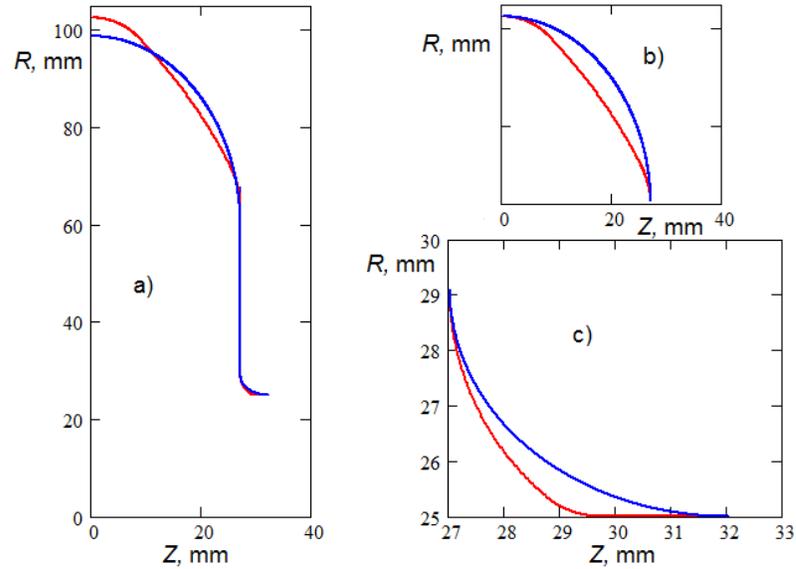

Figure 9. Substitution of the upper and the lower elliptic arcs (blue) with 3 circular arcs (red) in each case for decreasing $B_{pk}/E_{acc}$. The end points of the equatorial curves are made coinciding in the picture b), for comparison, the equatorial radius is increased for tuning to the same frequency, a).

## XI. GROUP VELOCITY AND CAVITY LENGTH LIMITATIONS

Coupling coefficient $k$ is related to the group velocity as follows:

$$k = 2\beta/(\theta \sin\theta),$$

where $\theta$ is the phase advance per cell, $\beta = v_{gr}/c$ is the group velocity normalized to speed of light. All the examples of cavities in Table III have $\beta > 0.01$. For a group velocity 0.01 and $\theta = 105°$ we have $k = 1.13 \cdot 10^{-2}$. We know that a SW $\pi$-mode structure can be tuned, it has $k_{SW} = 1.8 \cdot 10^{-2}$. To have the same error in the field, we must have a number of cells in a TW structure equal to

$$N_{TW} = 2(k_{TW}/k_{SW}) \cdot N_{SW}^2.$$

The length of this structure will be $L_s = N_{TW} \cdot L_c$, where $L_c$ is length of a cell, $L_c = \lambda\theta/(2\pi)$ Uniting all the above expressions, one can write

$$L_s = \frac{2 N_{SW}^2}{\pi k_{SW}} \cdot \frac{\beta\lambda}{\sin\theta}.$$

Even for $\sin\theta = 1$ the length of the structure will be 6.6 meters; it is much longer than can be permitted by technological limitations. Even a small aperture structure (Table IV) can be long enough having $\beta = 0.009$, close to 0.01. If the structure length is 2.4 m, we can remove the nonaccelerating intervals (Fig. 10) between the cavities ($\approx 40$ cm) and have a real-estate gradient 2.4 m/2 m $\approx 1.2$ times higher, e.g. 84 MeV/m instead of 70 MeV/m.

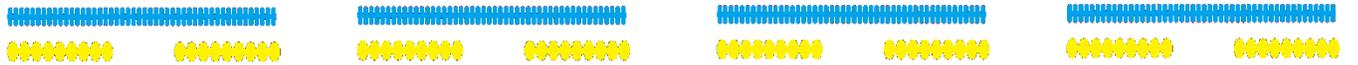

Figure 10. Removing nonaccelerating intervals between cavities: comparison with the TESLA cryomodule comprising 8 cavities.

## XII. DISPERSION CURVES

Dispersions of the first ten monopole modes for geometries from the first columns of Table V and Table VI are presented in Table VII and in Figs. 11 and 12. Coupling coefficients defined as $2(f_\pi - f_0)/(f_\pi + f_0)$ are presented Table VII and in Fig. 13.

One can see that coupling for both geometries is large enough and no trapped modes should be expected. The patterns of modes' positions are alike in both cases, so one can expect that other TW cavities discussed in this report will be free of trapped modes.

Frequencies of 0- and $\pi$-modes and coupling coefficients for the first dipole modes for same geometries as in Table VII are presented in Table VIII.

Table VII. Dispersion frequencies and coupling coefficients of the first 10 monopole modes for two cell geometries.

| Mode # | \multicolumn{9}{c|}{Geometry from the first column of Table V} | coupl. coeff. |
|---|---|---|---|---|---|---|---|---|---|---|
| | \multicolumn{9}{c|}{Phase advance} | |
| | 0 | 20 | 40 | 60 | 80 | 100 | 120 | 140 | 160 | 180 | |
| 1 | 1285.61 | 1286.36 | 1288.51 | 1291.80 | 1295.79 | 1300.00 | 1303.92 | 1307.09 | 1309.15 | 1309.86 | 0.0187 |
| 2 | 2867.02 | 2869.56 | 2876.96 | 2888.63 | 2903.56 | 2920.31 | 2936.99 | 2951.41 | 2961.28 | 2964.81 | 0.0335 |
| 3 | 3351.65 | 3350.56 | 3347.31 | 3342.05 | 3335.08 | 3326.92 | 3318.41 | 3310.74 | 3305.31 | 3303.34 | -0.0145 |
| 4 | 4295.64 | 4296.62 | 4299.58 | 4304.56 | 4311.56 | 4320.52 | 4331.08 | 4342.28 | 4351.83 | 4355.83 | 0.0139 |
| 5 | 4595.91 | 4592.96 | 4584.26 | 4570.32 | 4552.00 | 4530.64 | 4508.07 | 4486.85 | 4470.63 | 4464.29 | -0.0291 |
| 6 | 5645.95 | 5640.04 | 5623.31 | 5598.70 | 5570.38 | 5542.46 | 5518.17 | 5499.63 | 5488.10 | 5484.20 | -0.0291 |
| 7 | 5919.04 | 5906.59 | 5889.82 | 5876.29 | 5866.70 | 5860.35 | 5856.35 | 5853.96 | 5852.71 | 5852.32 | 0.0113 |
| 8 | 5938.87 | 5950.71 | 5966.17 | 5978.57 | 5987.66 | 5994.03 | 5998.35 | 6001.10 | 6002.63 | 6003.12 | 0.0108 |
| 9 | 6491.91 | 6500.35 | 6523.93 | 6558.25 | 6597.84 | 6637.58 | 6673.24 | 6701.46 | 6719.64 | 6725.93 | 0.0354 |
| 10 | 7158.02 | 7152.13 | 7137.59 | 7118.92 | 7098.97 | 7079.60 | 7062.25 | 7048.27 | 7039.04 | 7035.79 | -0.0172 |

| Mode # | \multicolumn{8}{c|}{Geometry from the first column of Table VI (transformed)} | coupl. coeff. |
|---|---|---|---|---|---|---|---|---|---|
| | \multicolumn{8}{c|}{Phase advance} | |
| | 0 | 22.5 | 45 | 67.5 | 90 | 112.5 | 135 | 157.5 | 180 | |
| 1 | 1284.29 | 1285.51 | 1288.97 | 1294.08 | 1300.00 | 1305.81 | 1310.66 | 1313.86 | 1314.97 | 0.0236 |
| 2 | 2905.47 | 2909.97 | 2923.01 | 2943.17 | 2968.08 | 2994.38 | 3017.98 | 3034.52 | 3040.49 | 0.0454 |
| 3 | 3599.25 | 3597.83 | 3593.62 | 3586.86 | 3578.06 | 3568.17 | 3558.76 | 3551.83 | 3549.25 | -0.0140 |
| 4 | 4407.03 | 4408.40 | 4412.50 | 4419.29 | 4428.58 | 4439.83 | 4451.67 | 4461.44 | 4465.38 | 0.0132 |
| 5 | 4742.75 | 4739.88 | 4731.36 | 4717.56 | 4699.34 | 4678.39 | 4657.65 | 4641.59 | 4635.37 | -0.0229 |
| 6 | 5824.25 | 5816.20 | 5792.21 | 5754.50 | 5709.32 | 5665.06 | 5628.91 | 5605.59 | 5597.57 | -0.0397 |
| 7 | 6139.73 | 6128.28 | 6102.62 | 6076.92 | 6058.01 | 6048.08 | 6039.22 | 6035.75 | 6034.70 | -0.0173 |
| 8 | 6477.78 | 6487.22 | 6508.50 | 6530.59 | 6548.10 | 6560.29 | 6568.02 | 6572.26 | 6573.61 | 0.0147 |
| 9 | 6834.77 | 6840.29 | 6855.49 | 6876.56 | 6898.66 | 6917.80 | 6931.90 | 6940.31 | 6943.09 | 0.0157 |
| 10 | 7435.65 | 7421.88 | 7403.75 | 7387.12 | 7372.99 | 7361.08 | 7353.97 | 7349.26 | 7347.70 | -0.0119 |

Table VIII. Frequencies and coupling coefficients for the first 10 dipole modes for two cell geometries.

| \multicolumn{4}{c}{Geometry from the first column of Table V} |
|---|---|---|---|
| Mode # | 0-mode frequency, MHz | $\pi$-mode frequency, MHz | coupling |
| 1 | 2896 | 2406 | 0.185 |
| 2 | 3744 | 3495 | 0.069 |
| 3 | 3908 | 3878 | 0.008 |
| 4 | 4981 | 4707 | 0.057 |
| 5 | 5149 | 5127 | 0.004 |
| 6 | 6232 | 5683 | 0.092 |
| 7 | 6452 | 6423 | 0.005 |
| 8 | 7447 | 6546 | 0.129 |
| 9 | 7618 | 7723 | 0.014 |
| 10 | 7808 | 7781 | 0.003 |

| \multicolumn{4}{c}{Geometry from the first column of Table VI (transformed)} |
|---|---|---|---|
| Mode # | 0-mode frequency, MHz | $\pi$-mode frequency, MHz | coupling |
| 1 | 3157 | 2510 | 0.228 |
| 2 | 3840 | 3641 | 0.053 |
| 3 | 4087 | 4031 | 0.014 |
| 4 | 5034 | 4815 | 0.044 |
| 5 | 5317 | 5262 | 0.010 |
| 6 | 6385 | 5995 | 0.063 |
| 7 | 6678 | 6629 | 0.007 |
| 8 | 7647 | 6831 | 0.113 |
| 9 | 8098 | 7998 | 0.012 |
| 10 | 8376 | 8060 | 0.038 |

## XIII. INFLUENCE OF FABRICATION ERRORS

Dimensions of the cells' elements in the Tables above are often given with an accuracy of micrometers. These are theoretical values which barely can be reached. So, we need to define possible deviations of these dimensions that do not compromise most important figures of merit. The value of $B$, the half-axis of the big ellipse, was already discussed: we changed it within several millimeters and had $E_{pk}/E_{acc}$ and $B_{pk}/E_{acc}$ changed within 1 %. Only slightly more sensitive are these parameters to deviation of $A$, another half-axis of the big ellipse, parallel to the axis of rotation. Half-axes of the small ellipse, $a$ and $b$, are much shorter than in the case of a π-mode SW structure (TESLA, e.g., has $a$ = 12 mm and $b$ = 19 mm). The sensitivity of the peak fields to the values of $a$ and $b$ is illustrated in Fig. 14. Contrary to the analytic functions, the derivative of the $E_{pk}/E_{acc}$ relative to $a$ or $b$ at its minimum is not zero but changes its value and sign at the point of optimum. So, this function is very sensitive to errors. This class of functions has a name of ravine functions: their minimum is like a surface of a creek's water at the bottom of a ravine with steep walls. The value of the function grows fast if one moves along the slope and slowly changes when one moves along the water flow. This property makes possible to change the cavity geometry without changing important parameters like $E_{pk}/E_{acc}$ if this change is done in a correct direction," along the creek".

For the case presented in Fig. 14, the growth of $E_{pk}/E_{acc}$ is 1 % when *a* increases by 0.580 mm, and this growth is also 1 % when *a* decreases by 0.058 mm, so the sensitivity differs 10 times. The value of $B_{pk}/E_{acc}$ depends linearly on *a* and changes by ±1 % when *a* changes by ±0.590 mm.

The growth of $E_{pk}/E_{acc}$ is 1 % for an increase of *b* by 0.190 mm or a decrease by 0.305 mm. Sensitivity of $B_{pk}/E_{acc}$ to the change of *b* is about 0.06 % per 1 mm.

Sensitivities of other geometries presented in the Table VI to fabrication errors are close to the values displayed here.

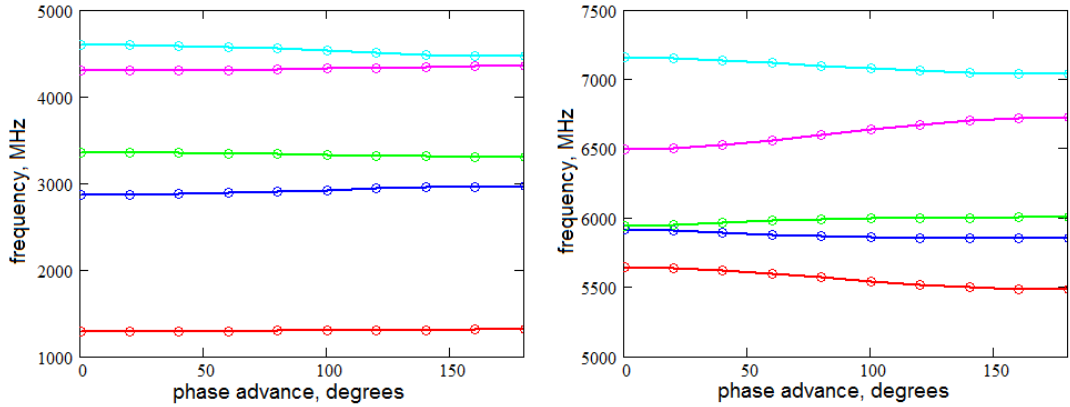

Figure 11. Dispersion curves for geometry from the first column of Table V.

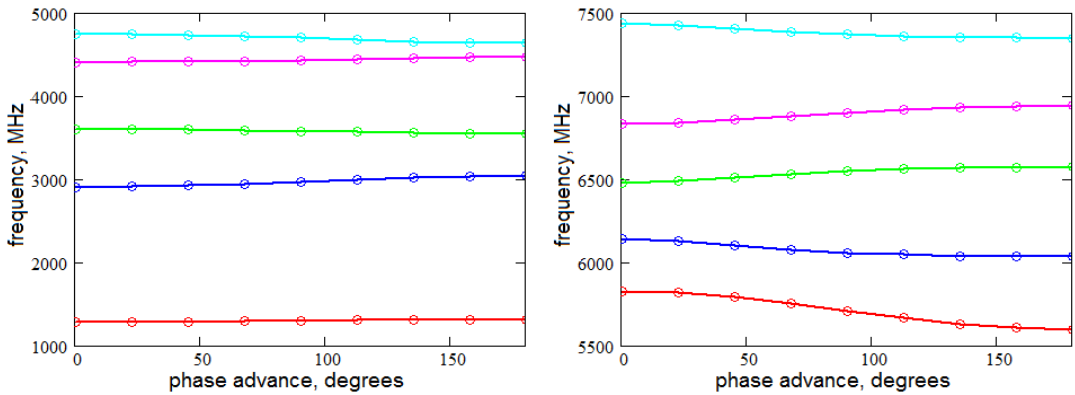

Figure 12. Dispersion curves for geometry from the first column of Table VI.

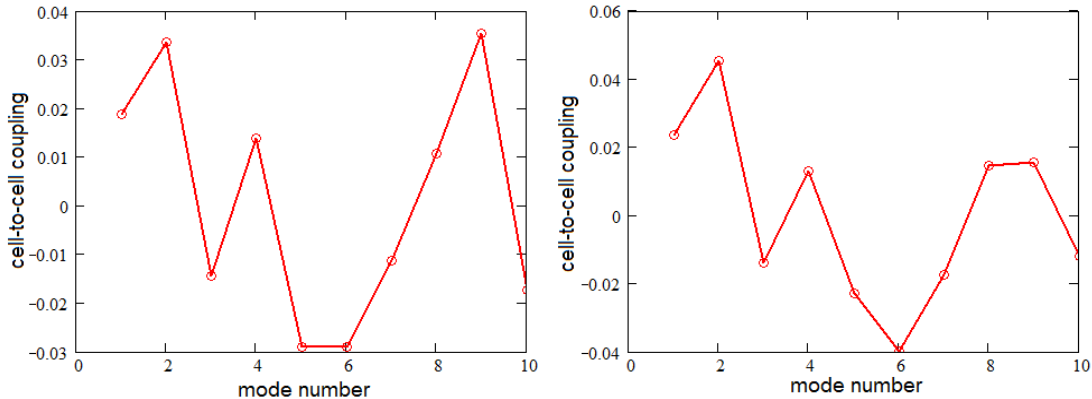

Figure 13. Coupling coefficients for the models presented in Fig. 11 (left) and 12.

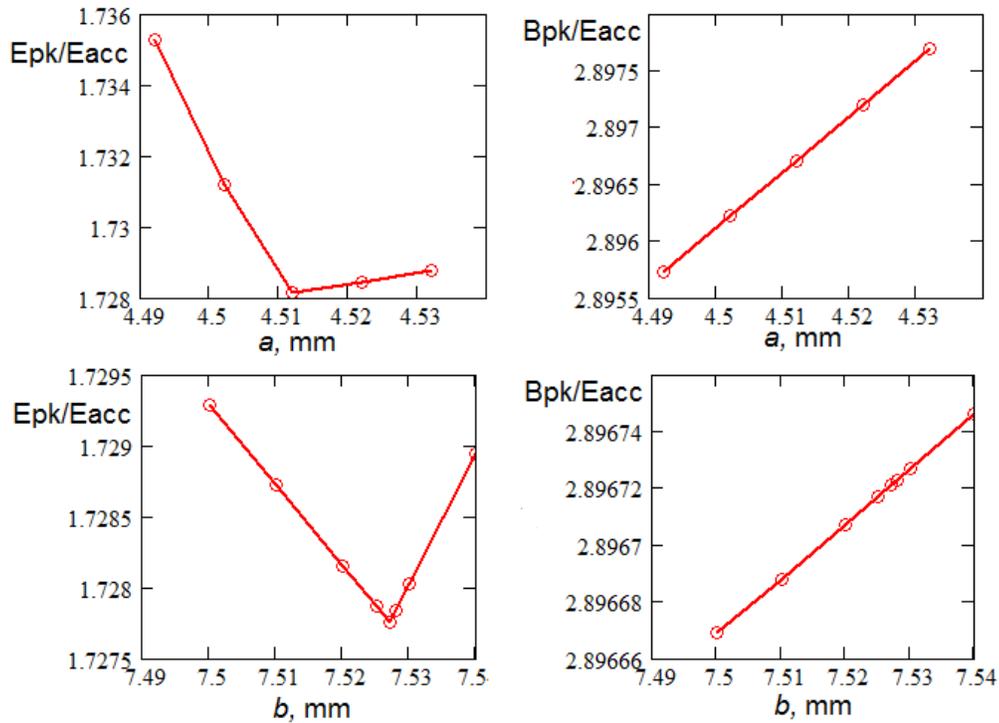

Figure 14. Dependencies of $E_{pk}/E_{acc}$ and $B_{pk}/E_{acc}$ on variations of lengths $a$ and $b$ for a cavity cell with $\theta = 90°$ and $B = 24$ mm from Table VI.

## XIV. LOSSES IN THE FEEDBACK WAVEGUIDE

It is supposed that a traveling wave structure is used in a resonant ring configuration [8–11]. Hence, a portion of the circulating power will be absorbed in the feedback waveguide and decrease the efficiency of the structure. In this section we estimate this effect.

The Q-factor of a structure is

$$Q = \frac{\omega_0 W}{P_{loss}},$$

where $\omega_0$ is the operating frequency, $W$ is the stored energy, and $P_{loss}$ is power loss in the structure, so

$$W = \frac{P_{loss} Q}{\omega_0}.$$

An energy per unit length is $W/L_s$, where $L_s$ is the structure length, and a flow of power through any cross-section of the structure is

$$P = \frac{W}{L_s} \cdot v_{gr} = \frac{P_{loss} Q}{L_s \omega_0} \cdot v_{gr},$$

where $v_{gr}$ is the group velocity. Here we assumed that the power density is the same in any cross-section, or power loss is much less than power circulating in the ring: $P_{loss} \ll P$.

Power loss can be expressed as

$$P_{loss} = \frac{V^2}{R},$$

$V$ is the cavity voltage, $R$ is the cavity shunt impedance. So, the power circulating in the ring is

$$P = \frac{V^2}{R} \cdot \frac{Q}{L_s \omega_0} \cdot v_{gr} = \frac{V^2}{R_{sh}/Q} \cdot \frac{v_{gr}}{L_s^2 \omega_0}.$$

Here $R_{sh}/Q$ is in Ohm/m, whereas $R$ is measured in Ohms. This is why $L_s$ appears squared.

Using $L_s = 1$ m, $V = 70$ MV, $R_{sh}/Q = 2000$ Ohm/m, $f = \omega_0/2\pi = 1.3 \cdot 10^9$ Hz, and $v_{gr} = 0.01c$, we calculate the power circulating in the ring $P = 900$ MW. Power loss in the whole structure is

$$P_{loss} = \frac{V^2}{(R_{sh}/Q) Q L_s} = 240 \text{ W}.$$

$Q$ is taken as $10^{10}$.

The attenuation coefficient of a rectangular waveguide for the $H_{10}$ wave is

$$\eta = \frac{2 k_r R_s}{b Z_0} \left[ \frac{1}{\sqrt{K}} \left( \frac{1}{2} + \frac{h}{d} \frac{f_{cr}^2}{f^2} \right) \right],$$

where $k_r$ is surface roughness coefficient (here, for simplicity $k_r = 1$), $R_s$ is the surface resistance, $Z_0$ is the impedance of free space, $K = 1 - f_{cr}^2/f^2$, $f_{cr} = c/(2d)$ is the critical frequency of the waveguide, $d$ and $h$ are the transverse dimensions (width and height) of the waveguide.

Taking dimensions of a standard WR-650 waveguide $d \times h = 165.1$ mm $\times$ 85.55 mm and surface resistance of the superconducting niobium $R_s = 27$ nOhm (this value corresponds to $Q = 10^{10}$ for the TESLA cavity) we have $\eta = 1.8 \cdot 10^{-9}$ 1/m. Attenuation of power follows the law: $P = P_0 \cdot \exp(-2\eta L)$. For the given values (waveguide length $L = 1$ m, $P_0 = 900$ MW) $\Delta P = P_0 - P = P_0[1 - \exp(-2\eta L)] = 3.6 \cdot 10^{-9} P_0 = 3.2$ W, so losses in the waveguide are about 1.3 % of the losses in the structure.

# XV. CONCLUSION

In this paper we presented results of modelling traveling wave structures aimed to keep both $B_{pk}$ and $E_{pk}$ values below limiting levels. A method of optimization is proposed that considers experimentally known limiting electric and magnetic surface fields. It is shown that a TW structure can have the accelerating gradient above 70 MV/m with the same critical magnetic field that the contemporary standing wave structures have. The other demonstrated benefit of TW structures is that their $R_{sh}/Q$ is about 2 times higher than for TESLA structure, that is equivalent to a factor of 2 higher $Q$ for reducing dynamic heat load. A multipactor suppression method is proposed: by sacrificing less than 1 % of the accelerating rate one can make the TW cavity multipactor-free. A group velocity for all simulated structures was calculated. The results show that cell-to-cell coupling is high enough to permit a very long cavity, so that the length will be limited only by fabrication considerations. An estimation of tolerances for fabrication the cavity cells is done. A design solution with the iris disc is proposed that can ease welding of the cells together and improve quality of the iris surface.

Many significant challenges remain on the path toward development of practical TW niobium structures. But the time scale for accomplishing this is several decades before the ILC 3 TeV upgrade is ready to launch. Below we list some of the challenges. There is an increased complexity due to doubling the number of cells (e.g., more welds). The cavity fabrication and surface processing fixtures and procedures must be modified and qualified. High circulating power in the feedback waveguide must be demonstrated. HOM damping must be studied. Preliminary results presented here show that the first 10 monopole modes up to 7 GHz show no trapping. At 3 times lower bunch charge for the ILC 3 TeV, HOM generation is much reduced over the ILC 0.5 and 1 TeV cases. The smaller aperture proposed in this work- means higher transverse wakes, but again the three times lower bunch charge helps. All the challenges are magnified if structures longer than 1 m are to be developed to further increase the gradients.

# XVI. ACKNOWLEDGMENT

The authors want to express a great gratitude to Sergey Belomestnykh for his support ad help.